# A STUDY ON THE EFFECT OF AGE, GENDER AND PARALYSIS ON sEMG SIGNALS

[1]Abhishek Jha, [2]Mrinal Sen
[1, 2] Department of Electronics Engineering, Indian School of Mines, Dhanbad, Jharkhand, India-826004.
Email: [1] abhishek.jha@ismu.ac.in

**Abstract**

Surface Electromyography (sEMG) is a technology to measure the bio-potentials across the muscles. The true prospective of this technology is yet to be explored. In this paper, a simple and economic construction of a sEMG sensor is proposed. These sensors are used to determine the differences in the Electromyography (EMG) signal patterns of different individuals. Signals of several volunteers from different age groups, gender and individual having paralysis have been obtained. The sEMG data acquisition is done using the soundcard of a computer, hence reducing the need of additional hardware. Finally, the data is used to analyse the relationship between electromyography and factors like age, gender and health condition i.e. paralysis.

## Introduction

It is observed that generally the muscular strength of males are more in compared to that of the females, also this strength decreases with the increase in the age of human. People with disability or paralysis show less muscular strength. Many authors have already proposed the linear relationship between EMG and force produced by a muscle [1-4]. Electromyography (EMG) can serve us to understand the reason behind these biological differences. Yet, very less study on the effect of age, gender and paralysis on the strength of the muscle using sEMG has been done.

sEMG signal is superimposed of many motor unit action potentials in muscle in time and space, which reflects functional status of nerve and muscle. EMG signal recovery using sEMG sensors is often difficult because the amplitude of sEMG signal of a healthy person is of range 10uV to 5000uV and lies in the frequency range 10Hz to 500 Hz [5]. Moreover it has very small SNR, the raw sEMG signal contains interference, also called hum, from

50Hz/60Hz AC power line sources [6]. Proper precautions were taken in order to minimize these interference noises. We have employed more than 20 test subjects of different age group in order to determine a general relationship between the force and EMG pattern of different individuals. Feature extraction, data processing and digital filtering of noise signals were done using custom made program in MATLAB.

## Construction

The basic requirement for the extraction of EMG signal from a muscle depends on the following factors; amplification, filtering and processing. Since the 50Hz interference induces a common mode signal which is stronger than sEMG, we therefore require a differential amplifier having high CMRR, around 100 [7]. Hence an instrumental amplifier AD620 is used for this purpose.

The preamplifier is using instrument amplifier AD620 to amplify sEMG signals in the first stage. AD620 has the merits of low power, high accuracy and low noise. The input offset voltage is 50uV max, input offset drift is 0.6 uV/ °C max and CMRR=120dB (G=10) [8]. The gain is calculated using (1).

$$G = 49.4k\Omega / Ro + 1 \qquad (1)$$

Here, Ro is the resistor between pin 1 and pin 8 of AD620 (Fig.1) and G is the gain of the amplifier.

G should be large enough because sEMG signals are very weak and prone to other noise. Again, if G is too large, it will make preamplifier get into saturation. By experimentation with different gain, we find the appropriate gain to be G=12 for this experiment. Hence the Ro becomes 4.25kΩ from (1).

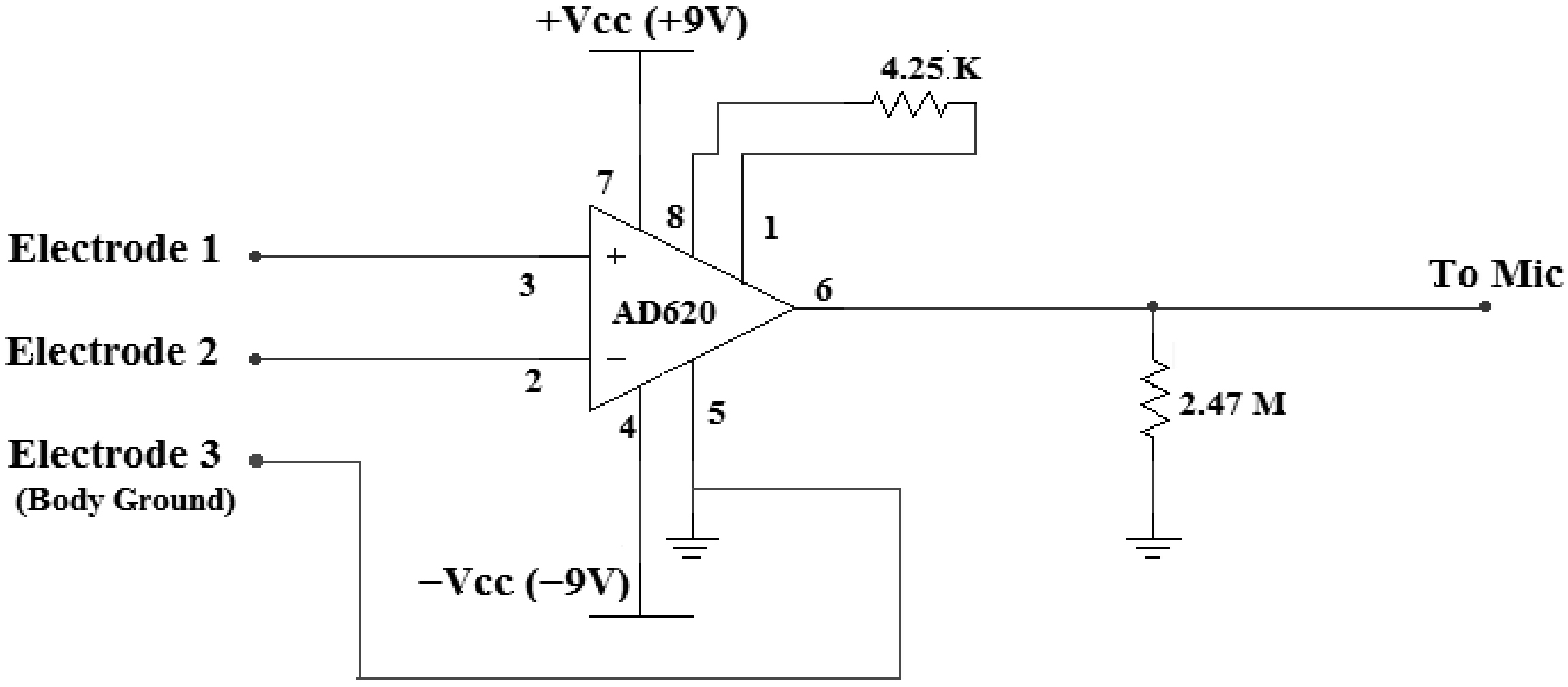

Fig.1: Circuit Diagram for sEMG sensor.

*Electrodes:*

In our experiment Dry Electrodes are used and the electrode 1 is separated by a distance of 3cm from electrode 2. Electrode 3 is attached to the body ground, usually the bony part, in this experiment it was the elbow joint.

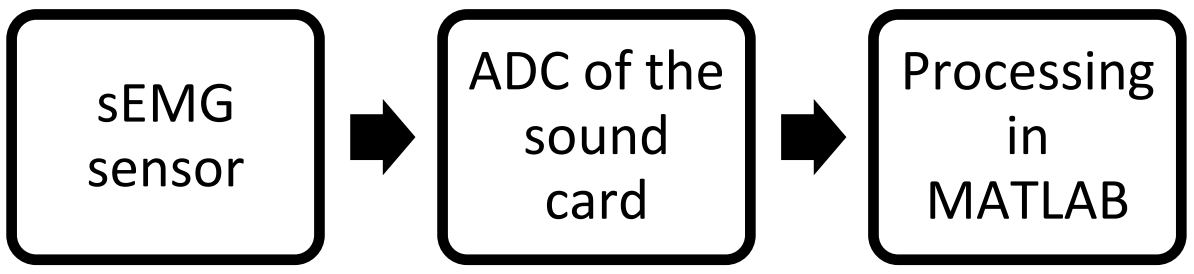

Fig.2 Block Diagram of the process.

*Processing:*

To process the output signal from the instrumental amplifier we passed it through the sound card of the computer, which is a 16 bit ADC (Fig.2). The Digital Signal is then processed using custom made MATLAB program.

*Filtering:*

The raw sEMG signal contains power line 50Hz/60Hz interference [6]. We therefore digitally diminished the 50Hz frequency using MATLAB. EMG signals lies in the range of 10 Hz to 500Hz. Hence, we used another digital bandpass filter in MATLAB, to get the desired frequency range.

**Method**

*Subjects:*

In total 36 volunteers were employed belonging to age groups 16 to 25, 35 to 45 and 55 to 65. The signals were also retrieved from a Volunteer who had paralysis in right hand. The test was done when the volunteer was recovering from paralysis.

| No. of Volunteers | Age Group | Gender | Normal/Paralytic |
|---:|---:|---:|---:|
| 10 | 16-25 | M | Normal |
| 10 | 16-25 | F | Normal |
| 5 | 35-45 | M | Normal |
| 5 | 55-65 | M | Normal |
| 5 | 55-65 | F | Normal |
| 1 | 35-45 | F | Half-Paralytic |

Table.1: Details of volunteers.

*Protocol:*

Each volunteer's age, gender, medical history, weight and general information like whether they do exercises of the particular muscle or not was noted down; in this experiment the target muscle is Flexor Carpi Radialis (Fig.3). Since the readings were taken in two positions i.e. relaxed and fully excited, the volunteers were instructed and explained to put their arm in two positions; i.e. relaxed position and excited position. The skin above the target muscle was cleaned and the hair was removed from that area in order to decrease the electrode skin impedance. Electrode 3 was attached to the elbow joint and electrode 1 and electrode 2 were attached just above the target muscle while maintaining the 3cm separation between both the electrodes. Few observations in excited positions were retaken because of some error in data due to dryness in the skin of some individuals. In such cases the individuals were given sufficient amount of time before retaking the data so that the excited muscle can be brought down again in the relaxed position.

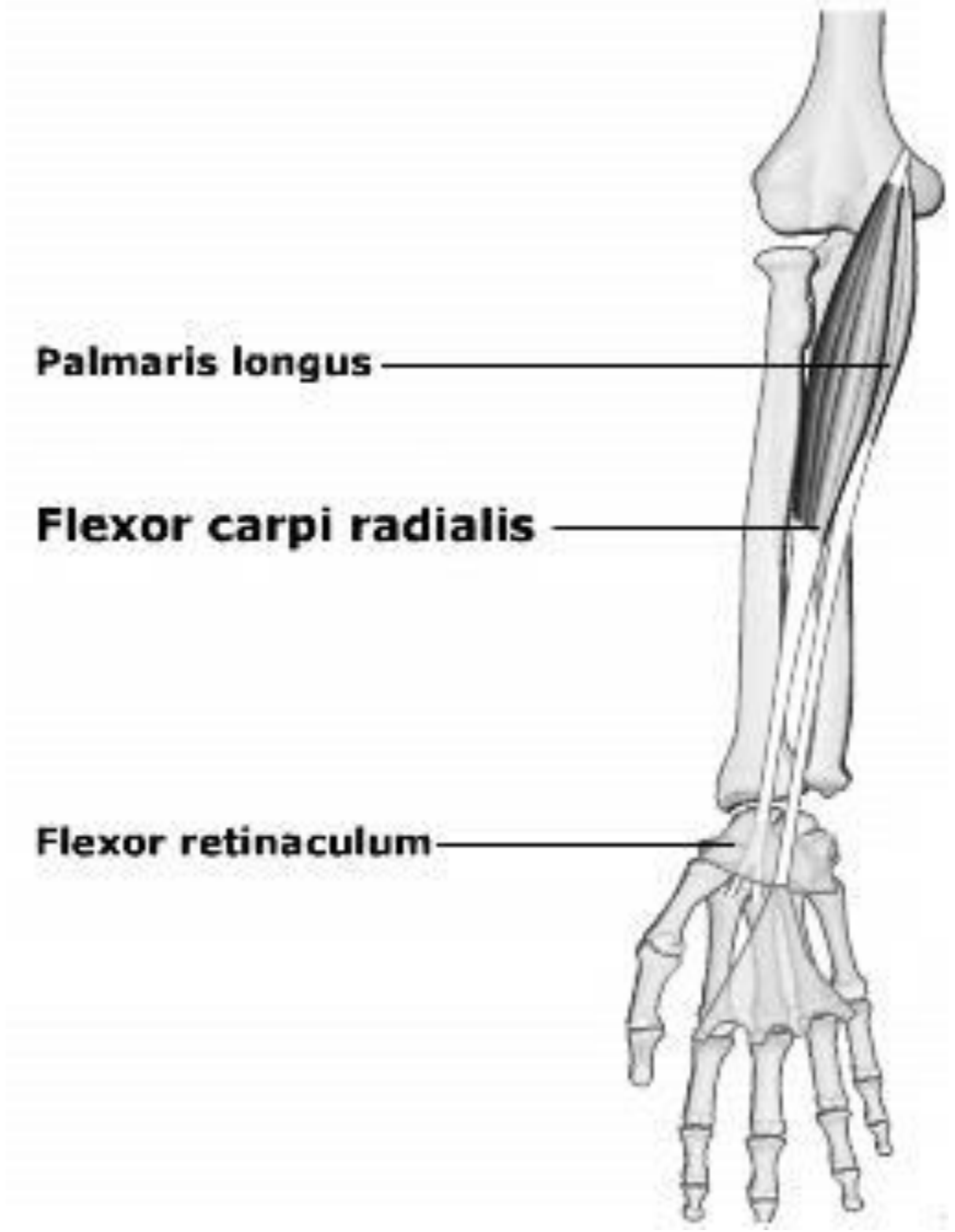

Fig.3: Flexor carpi radialis

*Data Collection:*

Each volunteer was instructed to first keep his right arm muscle, Flexor Carpi Radialis, in the fully relaxed condition 20 samples were taken at an interval of 500ms. The input data was in time domain, therefore those were converted into the frequency domain by taking their Fast Fourier Transform (FFT), and unwanted frequencies were omitted. Frequency response of the sample consisted of frequencies separated by 3.9Hz. Then keeping the sensors at the same place on the arm, 20 samples were taken while the volunteers were asked to keep their muscle (Flexor Carpi Radialis) in fully excited position by tightening their fist. Similarly, the data was taken for the left hand in the two positions.

*Data Analysis:*

Data in fully excited and fully relaxed condition were compared for each volunteer. Data of same group were averaged to get a general relationship between force and sEMG value of individuals, as well as the data of one age group were compared with that of another to relate the effect of age on the sEMG signal. Similarly, the data of female of each age group is compared to that of the male of corresponding age group. Data of people having paralysis were compared with the data of normal individual of the corresponding age group. Also the data of the left arm of each individual was compared to that of the right arm.

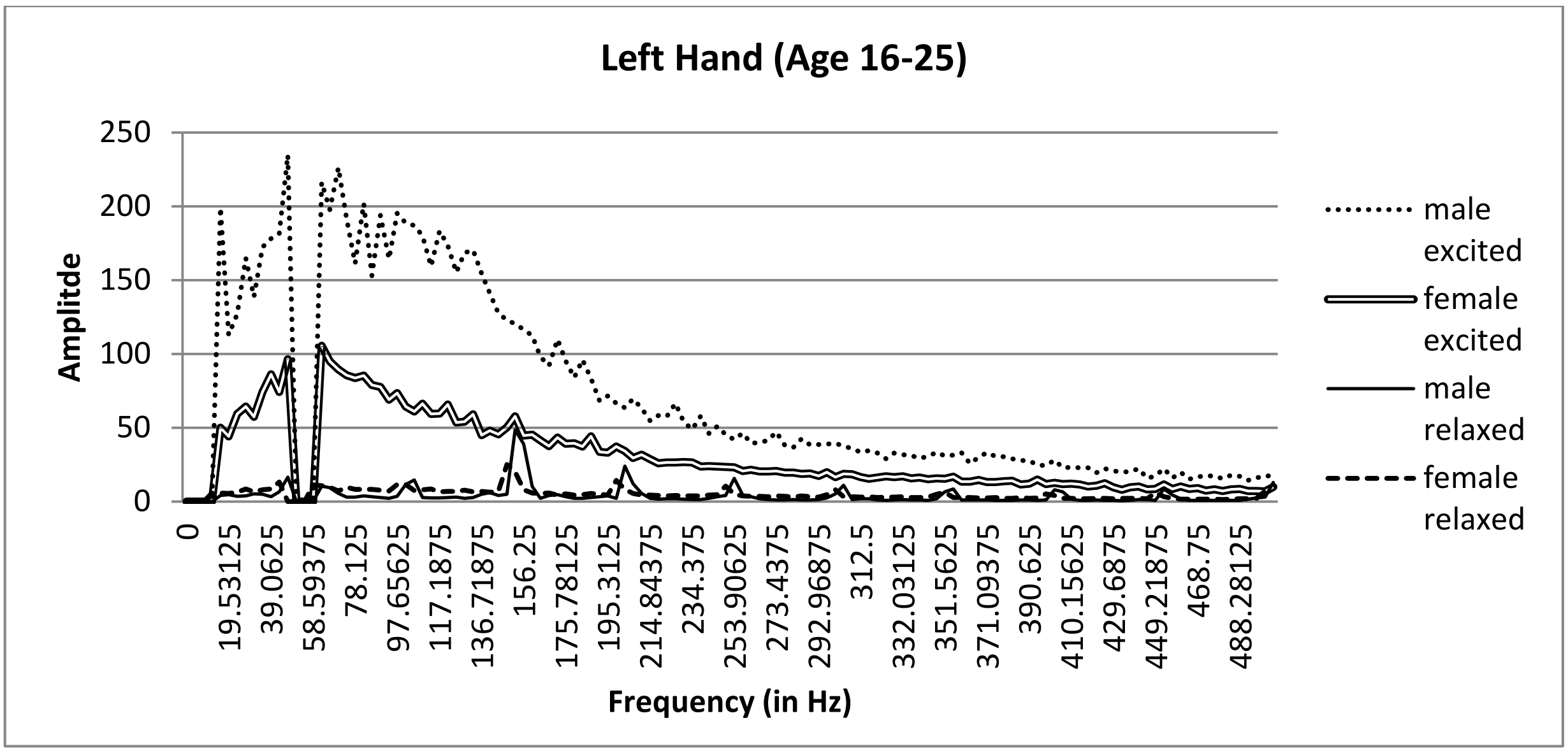

Graph.1: Left hand sEMG data for age group 16-25.

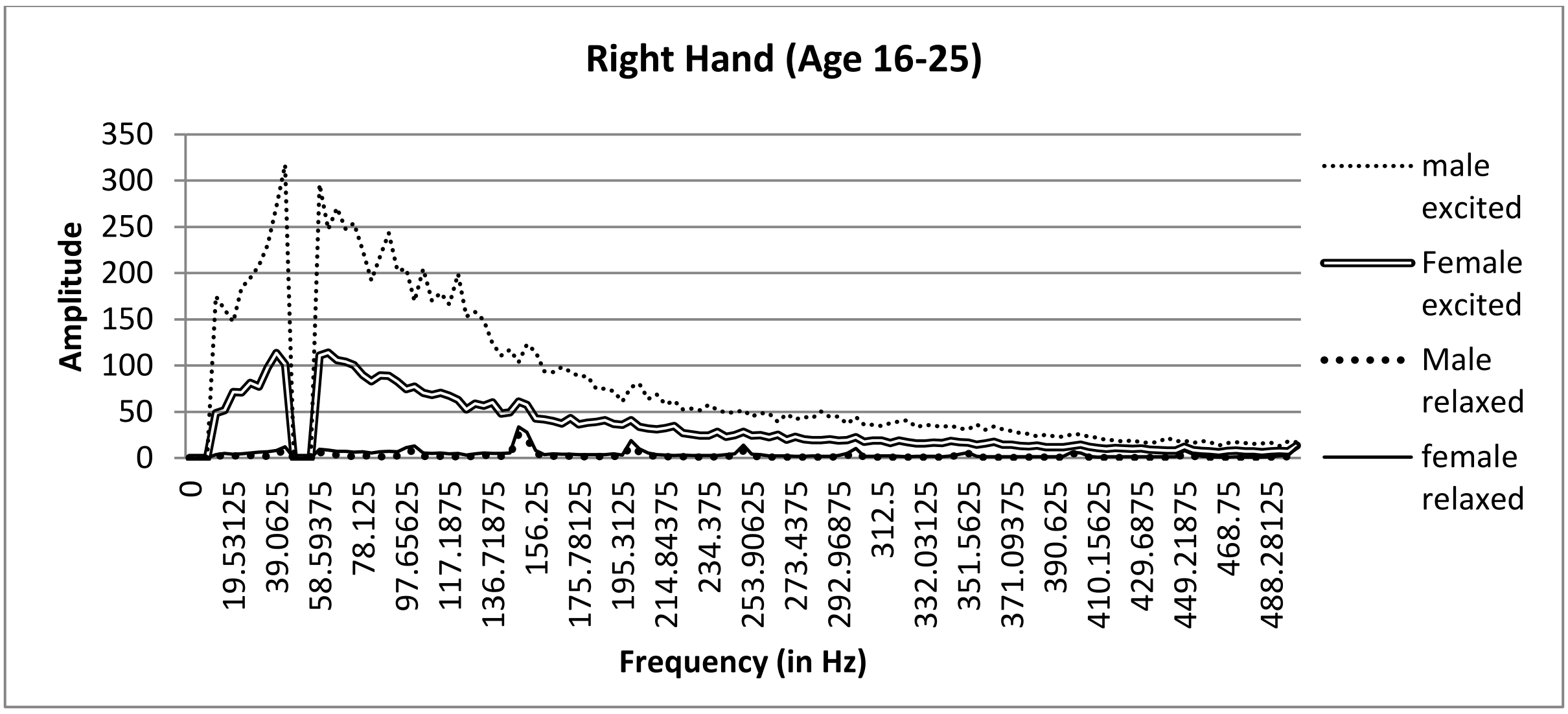

Graph.2: Right hand sEMG data for age group 16-25.

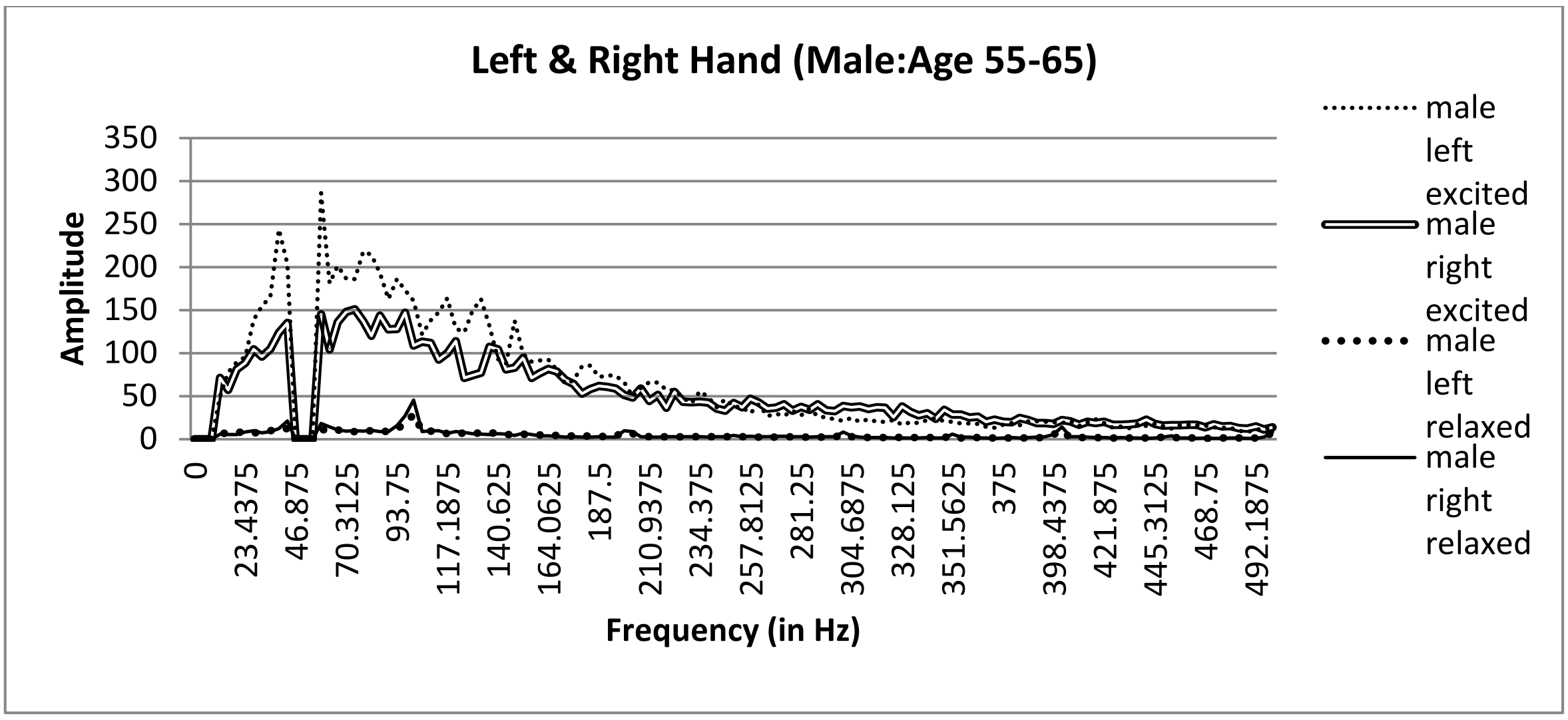

Graph.3: Right and Left hand sEMG data for male of age group 55-65.

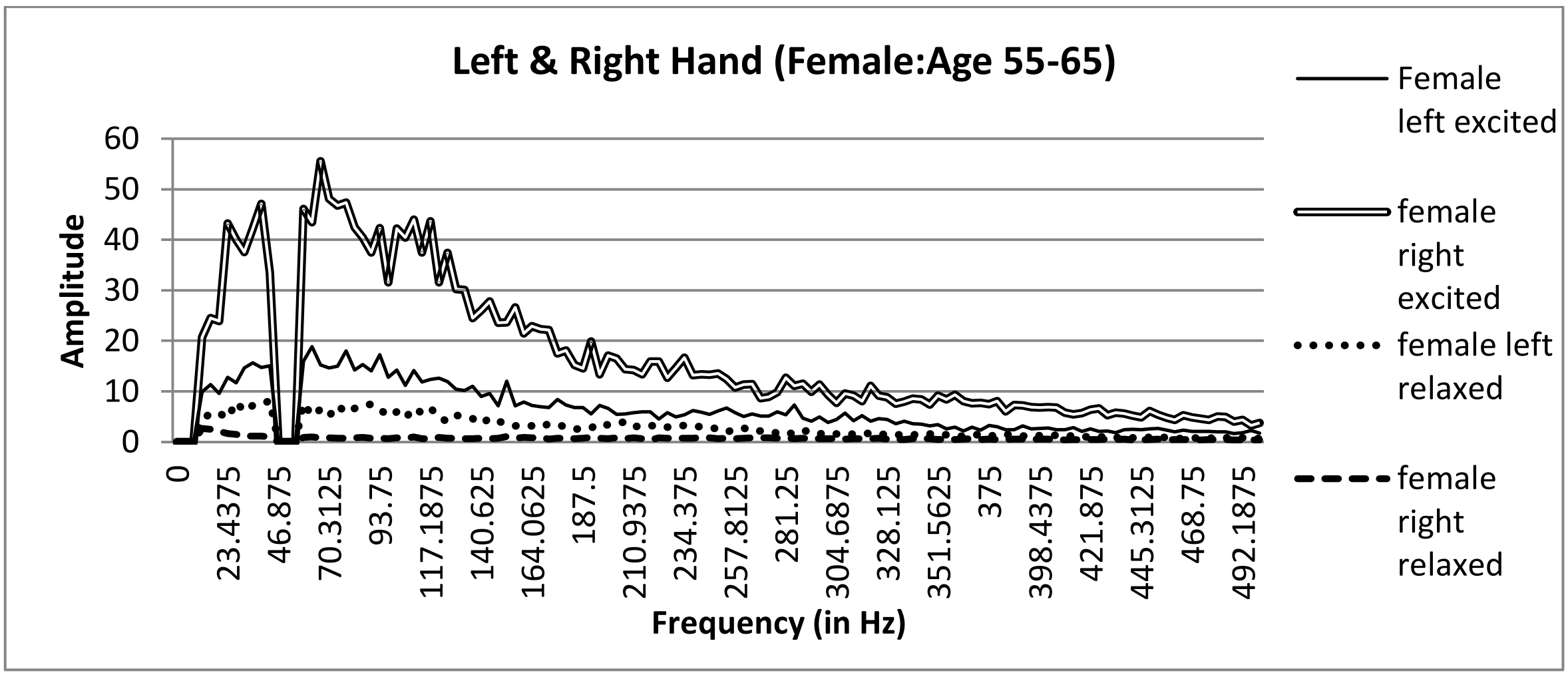

Graph.4: Right and Left hand sEMG data for female of age group 55-65.

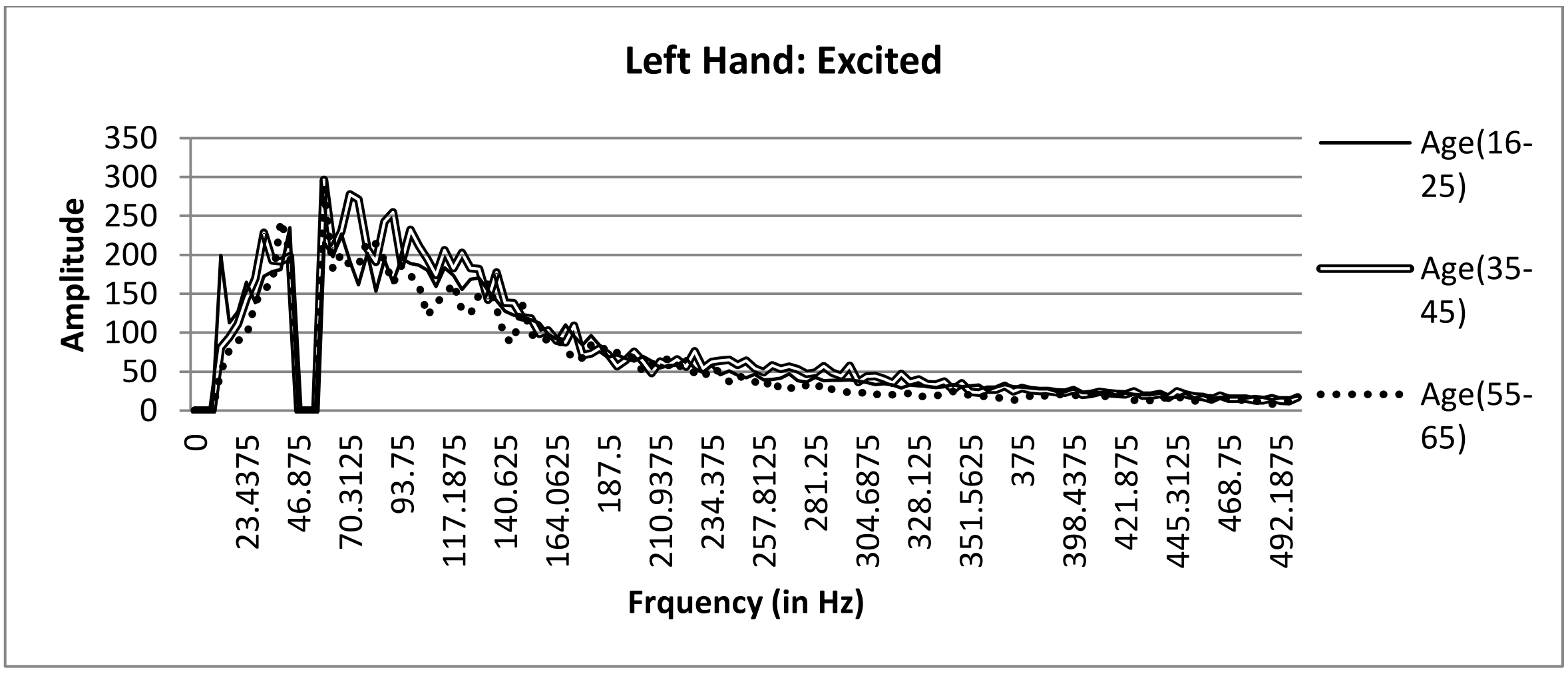

Graph.5: Comparison between the left hand sEMG data of different age groups at fully excited position.

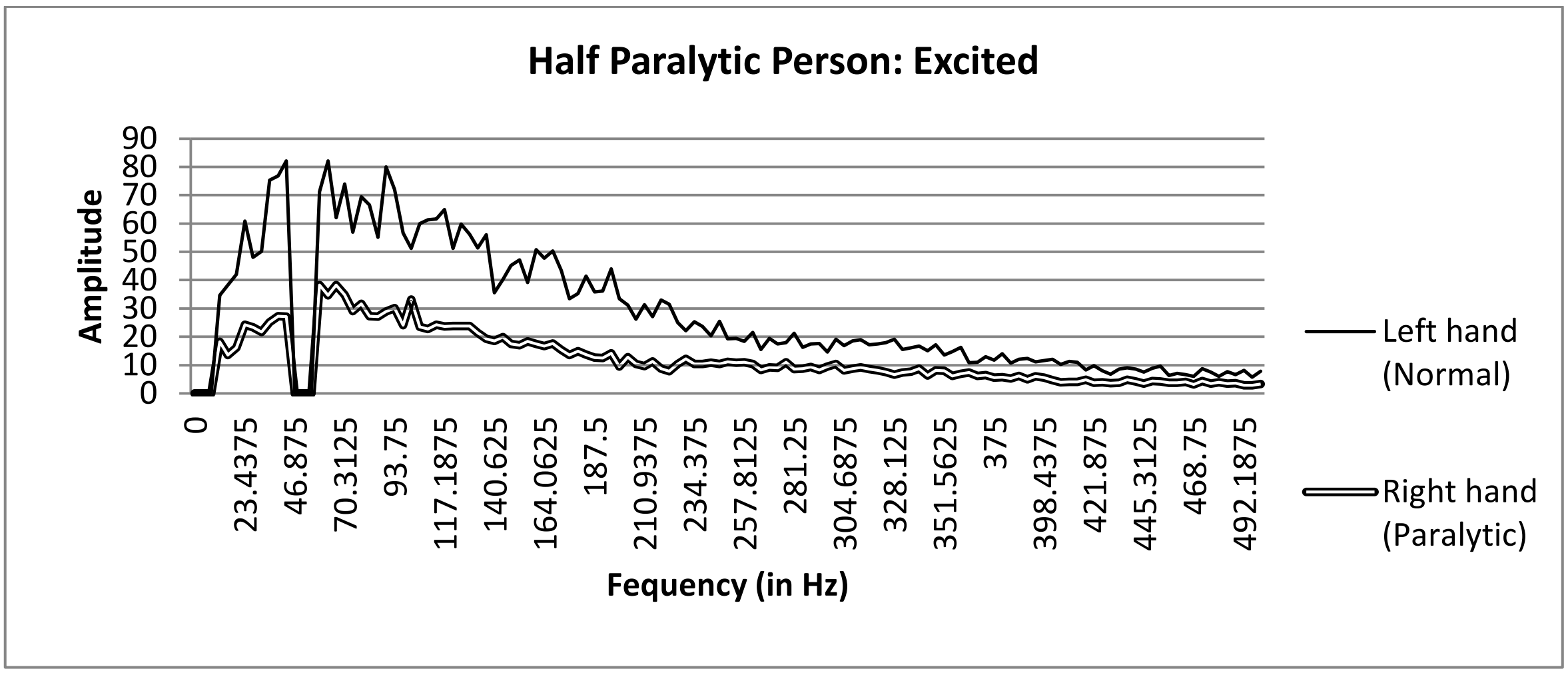

Graph.6: Comparison between the sEMG data of the normal hand to the paralytic hand of a half paralytic individual of age group 35-45.

## Result

It has been observed that the right hand’s sEMG signal in all the age group is more in comparison to that of the left hand(Graph.1 and Graph.2), hence justifying the conventional observation that a right handed person’s right hand is stronger in comparison to his left hand. The sEMG data of female is found to be less than in comparison to the male, in all age groups (Graph.1 to Graph.4), hence justifying the conventional belief that the muscular strength of male is more than female. Also it can be observed that the sEMG signal’s strength increases from age group (16-25) to age group (35-45) and then decreases for age group (55-65) (Graph.5), hence it shows that the muscle strength increases upto adult age and then decrease with increase in age. It has been observed that the paralytic hand shows less amplitude sEMG signals in comparison to the normal hand of the same individual(Graph.6) as well as in comparison to the normal human beings.

## Conclusion

A simple and economic sEMG sensor has been made and the data acquisition has been done using the soundcard of a computer reducing the hardware requirement, signals has been processed in the MATLAB software using custom made program. The conventional belief of the physiological difference between different individual on the basis of age, gender and health condition like paralysis has been justified using sEMG data. It has also been justified that the right handed individual’s left hand comparatively shows less strength than right hand. These data can be used as a reference to heal/cure the people with muscular incapability. The work can be extended by creating the database of major muscle of human body, which would serve for the purpose of creating prosthetic limbs for people with paralysis or amputees.